\documentclass[a4paper,10pt]{article}

\usepackage{amsmath,amssymb,bbm,mathtools}
\usepackage{ascmac}
\usepackage{cancel}
\usepackage{xcolor} 
\usepackage{graphicx} 
\newdimen\tbaselineshift 
\usepackage[bookmarks=true,bookmarksnumbered=true,bookmarkstype=toc]{hyperref} 
\hypersetup{
pdfauthor={Tetsuji KIMURA},
colorlinks={true},
linkcolor={blue},
urlcolor={blue},
filecolor={blue},
citecolor={blue}
}
\definecolor{refkey}{rgb}{0.40, 0.55, 0.55}
\definecolor{labelkey}{rgb}{0.40, 0.55, 0.55}
\usepackage{ulem}

 \font\f=cmr10
 \pdffontexpand\f 30 20 10 autoexpand
\definecolor{darkblue}{HTML}{00008B}
\definecolor{brown}{HTML}{A52A2A}
\definecolor{darkcyan}{HTML}{008B8B}
\definecolor{darkgreen}{HTML}{006400}
\definecolor{darkolivegreen}{HTML}{556B2F}
\definecolor{darkseagreen}{HTML}{8FBC8B}
\definecolor{deeppink}{HTML}{FF1493}
\definecolor{deepskyblue}{HTML}{00BFFF}
\definecolor{dodgerblue}{HTML}{1E90FF}
\definecolor{gray}{HTML}{808080}
\definecolor{hotpink}{HTML}{FF69B4}
\definecolor{lightblue}{HTML}{ADD8E6}
\definecolor{lightcyan}{HTML}{E0FFFF}
\definecolor{lightgray}{HTML}{D3D3D3}
\definecolor{lightgrey}{HTML}{D3D3D3}
\definecolor{lightpink}{HTML}{FFB6C1}
\definecolor{navy}{HTML}{000080}
\definecolor{orange}{HTML}{FFA500}
\definecolor{orchid}{HTML}{DA70D6}
\definecolor{pink}{HTML}{FFC0CB}
\definecolor{purple}{HTML}{800080}
\definecolor{royalblue}{HTML}{4169E1}
\definecolor{seagreen}{HTML}{2E8B57}
\definecolor{skyblue}{HTML}{87CEEB}
\definecolor{steelblue}{HTML}{4682B4}

\makeatletter
\def\tbcaption{\def\@captype{table}\caption}
\def\figcaption{\def\@captype{figure}\caption}
\makeatother

\newcounter{Enumerate}

\DeclareFontFamily{U}{rsf}{}
\DeclareFontShape{U}{rsf}{m}{n}{
  <5> <6> rsfs5 <7> <8> <9> rsfs7 <10-> rsfs10}{}
\DeclareMathAlphabet\Scr{U}{rsf}{m}{n}

\usepackage[mathscr]{eucal}

\newcommand{\half}{\frac{1}{2}}

\newcommand{\LS}{\ \ \ \ \ \ \ \ \ \ }
\newcommand{\ls}{\ \ \ \ \ }

\newcommand{\bsubeq}{\begin{subequations}}
\newcommand{\esubeq}{\end{subequations}}

\newcommand{\nn}{\nonumber}

\newcommand{\A}{\mathscr{A}}
\newcommand{\B}{\mathscr{B}}

\newcommand{\T}{{\rm T}}
\newcommand{\X}{\mathscr{X}}
\newcommand{\Y}{\mathscr{Y}}
\newcommand{\Z}{\mathscr{Z}}

\renewcommand{\d}{{\rm d}}
\newcommand{\e}{{\rm e}}

\newcommand{\slb}{\scalebox}

\begin{document}
\allowdisplaybreaks{

\thispagestyle{empty}


\begin{flushright}
TIT/HEP-658 
\end{flushright}

\vspace{5mm}

\begin{center}
\slb{1.5}{Explicit Description of the Zassenhaus Formula}
\end{center}

\begin{center}
  {\renewcommand{\arraystretch}{1.2}
\begin{tabular}{c}
\slb{1.2}{Tetsuji {\sc Kimura}} \vphantom{$\Bigg|$}
\\
\slb{.9}{\renewcommand{\arraystretch}{1.0}
\begin{tabular}{c}
{\sl Research and Education Center for Natural Sciences, Keio University}
\\
{\sl Hiyoshi 4-1-1, Yokohama, Kanagawa 223-8521, JAPAN} 
\end{tabular}
}
\\
\slb{.85}{and}
\\
\slb{.9}{\renewcommand{\arraystretch}{1.0}
\begin{tabular}{l}
{\sl
Department of Physics,
Tokyo Institute of Technology, Tokyo 152-8551, JAPAN}
\end{tabular}
}
\\
\slb{0.9}{\tt tetsuji.kimura \_at\_ keio.jp}
\end{tabular}
}
\end{center}

\begin{abstract}
We explicitly describe an expansion of $\e^{A+B}$ as an infinite sum of the products of $B$ multiplied by the exponential function of $A$.
This is the explicit description of the Zassenhaus formula.
We also express the Baker-Campbell-Hausdorff formula in a different manner.
\end{abstract}

\subsection*{Introduction}

In various topics in physics and mathematics,
we often have to expand the exponential function of two operators $A$ and $B$ such as $\e^{A+B}$ in a certain situation (for instance, \cite{Dyson-series} and \cite{Kimura:2016irk}).
An expansion is described as the Zassenhaus formula 
(\cite{Zassenhaus} and references therein):
\bsubeq \label{Zassenhaus-formula}
\begin{align}
\e^{t (A + B)}
\ &= \ 
\e^{t A} \e^{t B} \prod_{n=2}^{\infty} \e^{t^n Z_n (A,B)}
\, , \\
Z_n \ &= \ 
\frac{1}{n!} \Big\{ \frac{\d^n}{\d t^n} (\e^{- t^{n-1} Z_{n-1}} \cdots \e^{- t^2 Z_2} \e^{-t B} \e^{-t A} \e^{t (A + B)}) \Big\}_{t = 0} 
\, .
\end{align}
\esubeq
Its transposed version is also given as
\bsubeq \label{Zassenhaus-2}
\begin{align}
\e^{\beta (A + B)}
\ &= \ 
\Big( \prod_{\infty}^{n=2} \e^{\beta^n \Z_n} \Big) \e^{\beta B} \e^{\beta A}
\, , \\
\Z_n \ &= \ 
\frac{1}{n!} \Big\{ \frac{\d^n}{\d \beta^n} (\e^{\beta (A + B)} \e^{-\beta A} \e^{-\beta B} \e^{- \beta^2 \Z_2} \cdots \e^{-\beta^{n-1} \Z_{n-1}}) \Big\}_{\beta = 0}
\, . 
\end{align}
\esubeq
Unfortunately, however, the above two expressions are rather complicated 
because we sequentially obtain the explicit expression of higher order terms in the operators $A$ and $B$. 
In this paper, we will obtain a new description of the Zassenhaus formula
in which all of the higher order terms are explicitly expressed.

\subsection*{Derivation}

First of all, we expand $(A+B)^n$ and move all the operator $A$ to the right in each term, and define the following expression:
\begin{align}
(A + B)^n
\ &\equiv \ 
\sum_{m=0}^n \frac{n!}{m! (n-m)!} \, X_{m} \, A^{n-m}
\, , \label{A+B^n}
\end{align}
where $X_m$ are polynomials involving $B^l$, commutators $[A,[A,\cdots[A,B]]]$, and their multiplications.
By using $X_m$, we obtain the exponential function of $A+B$ as the following form:
\begin{align}
\e^{A + B}
\ &= \ 
\sum_{n=0}^{\infty} \frac{1}{n!} (A + B)^n
\ = \ 
\sum_{n=0}^{\infty} \sum_{m=0}^n \frac{1}{m!(n-m)!} X_{m} \, A^{n-m}
\ = \ 
\Big( \sum_{m=0}^{\infty} \frac{1}{m!} X_m \Big) \, \e^{A}
\, . \label{resum} 
\end{align}
There exists a recursion relation among $X_m$ in such a way that
\begin{align}
X_{m+1}
\ &= \ 
\mathscr{L}_{A} X_{m}
+ B X_{m} 
\, , \ls
X_0 \ \equiv \ 1
\, , \ls
X_1 \ = \ B 
\, , \label{Xk_Xk-1} 
\end{align}
where $\mathscr{L}_{A} {\cal O}$ is the commutator between $A$ and a certain operator ${\cal O}$ such as $\mathscr{L}_{\A} {\cal O} = [A, {\cal O}]$. 
It is easy to derive (\ref{Xk_Xk-1}) when we compute $(A+B)^{n+1}$ as the product of $(A+B)$ and $(A+B)^n$ in terms of (\ref{A+B^n}).
Let us evaluate the relation (\ref{Xk_Xk-1}).
It is convenient to express $X_m$ as the sum of new polynomials $X_{m,p}$:
\begin{align}
X_m \ &= \ 
\sum_{p=1}^m X_{m,p}
\, , \label{Xm-Xmp}
\end{align}
where $p$ denotes the power of $B$ in $X_{m,p}$, whose examples can be seen in the appendix.
Substituting the expression (\ref{Xm-Xmp}) into the relation (\ref{Xk_Xk-1}), we find three recursion relations
\bsubeq \label{re:Xm-gen}
\begin{alignat}{2}
X_{m+1,1} \ &= \ 
\mathscr{L}_{A} X_{m,1}
\, , \label{re:X_m1-gen} \\
X_{m+1,m+1} \ &= \ 
B X_{m,m}
\, , \label{re:X_mm-gen} \\
X_{m+1,p} \ &= \ 
\mathscr{L}_{A} X_{m,p}
+ B X_{m,p-1}
\, , &\ls
&m \geq p 
\, . \label{re:Xmp-gen}
\end{alignat}
\esubeq
We immediately obtain the solutions of these relations (the proof is exhibited in the appendix):  
\bsubeq \label{sol:Xm-gen}
\begin{align}
X_{m,1}
\ &= \ 
(\mathscr{L}_{A})^{m-1} X_{1,1}
\ = \ 
(\mathscr{L}_{A})^{m-1} B
\ \equiv \ 
\B'_m
\, , \label{sol:Xm1-gen} \\
X_{m,m}
\ &= \ 
B^{m-1} X_{1,1}
\ = \ 
B^m
\, , \label{sol:Xmm-gen} \\
X_{m,p}
\ &= \
\sum_{k=1}^{m-p+1} 
\frac{(m-1)!}{(k-1)! (m-k)!}
\, X_{m-k,p-1} \B'_k
\, . \label{sol:Xmp-gen}
\end{align}
\esubeq
Here we introduced the terminology $\B'_m$ defined the above.
The solution (\ref{sol:Xmp-gen}) can be described in an explicit way 
if we iteratively use (\ref{sol:Xmp-gen}) until we reach
$X_{m-(k_1 + \ldots + k_{p-1}),1} = \B'_{m-(k_1 + \ldots + k_{p-1})}$ 
given by (\ref{sol:Xm1-gen}).
Hence we obtain
\begin{align}
X_{m,p}
\ &= \ 
\sum_{k_1=1}^{m-p+1} \sum_{k_2=1}^{m-k_1 -p+ 2} 
\cdots
\sum_{k_{p-1}=1}^{m-(k_1 + \ldots + k_{p-2})-1}
\frac{m! \cdot k_1 k_2 \cdots k_{p-1}}{m (m-k_1) (m-k_1-k_2) \cdots (m-(k_1 + \ldots + k_{p-2}))}  
\nn \\
\ & \LS \times 
\B_{m-(k_1 + \ldots + k_{p-1})} \B_{k_{p-1}} \cdots \B_{k_2} \B_{k_1}
\, . \label{sol:Xmp-gen2}
\end{align}
For simplicity, we further introduced the description $\B_m \equiv \frac{1}{m!} \B'_m$.
Applying (\ref{sol:Xmp-gen2}) to (\ref{A+B^n}) and (\ref{Xm-Xmp}), 
we obtain the explicit expansion of $\e^{A+B}$ in terms of the products of $\B_m$:
\begin{align}
\e^{A + B}
\ &= \ 
\Big( \sum_{m=0}^{\infty} \frac{1}{m!} X_m \Big) \, \e^{A}
\ = \ 
\Big( 1 + \sum_{m=1}^{\infty} \sum_{p=1}^m \frac{1}{m!} X_{m,p} \Big) \, \e^{A}
\nn \\
\ &= \ 
\Big\{
1 + \sum_{m=1}^{\infty} \B_m
+ \sum_{m=1}^{\infty} \sum_{k_1=1}^{m-1} \frac{k_1}{m} \, \B_{m-k_1} \B_{k_1}
+ \sum_{m=1}^{\infty} \sum_{k_1=1}^{m-2} \sum_{k_2=1}^{m-k_1-1} \frac{k_1 k_2}{m (m-k_1)} \, \B_{m-k_1-k_2} \B_{k_2} \B_{k_1}
\nn \\
\ &\LS 
+ \sum_{m=1}^{\infty} \sum_{k_1=1}^{m-3} \sum_{k_2=1}^{m-k_1-2} \sum_{k_3=1}^{m-k_1-k_2-1} \frac{k_1 k_2 k_3}{m (m-k_1) (m-k_1-k_2)} \, \B_{m-k_1-k_2-k_3} \B_{k_3} \B_{k_2} \B_{k_1}
\nn \\
\ &\LS 
+ \ldots
\Big\} \, \e^{A}
\, . \label{explicit-Xmp-gen}
\end{align}
Relabeling $k_i$ and $m - (k_1 + \ldots + k_{p-1})$ to $n_i$ and $n_p$ respectively, we obtain the final form
\begin{align}
\e^{A + B}
\ &= \ 
\Big\{
1 + \sum_{p=1}^{\infty} \sum_{n_1, \ldots, n_p=1}^{\infty} 
\frac{n_p \cdots n_1}{n_p (n_p + n_{p-1}) \cdots (n_p + \ldots + n_1)} \, \B_{n_p} \cdots \B_{n_1}
\Big\} \, \e^{A}
\, . \label{A+B:genB-2} 
\end{align}
We have a comment that each $k_i$, as well as the new label $n_i$, is unbounded from above because $m$ goes to infinity.
It turns out that (\ref{A+B:genB-2}) is the 
the explicit description
of the Zassenhaus formula (\ref{Zassenhaus-2}) without using the functions $\Z_n$. 
We understand that the exponential form $\e^B$ in the right-hand side of (\ref{Zassenhaus-2}) can be obtained from (\ref{A+B:genB-2}) when we extract the terms of the products only of $\B_{1} = B$.
However, it is hard to extract $\e^{\Z_m}$ of arbitrary $m$ from (\ref{A+B:genB-2}).
Because of the iterative definition of $\Z_m$ in (\ref{Zassenhaus-2}),
we have to obtain the explicit expression of all $\Z_l$ where $l \leq m-1$ beforehand to determine $\Z_m$.

Furthermore, if we transpose (\ref{A+B:genB-2}) and rename $A^{\T}$ and $B^{\T}$ to $A$ and $B$, we obtain 
\begin{align}
\e^{A + B}
\ &= \ 
\e^{A} \, \Big\{
1 + \sum_{p=1}^{\infty}
\sum_{n_1, \ldots, n_p=1}^{\infty} 
\frac{(-1)^{(n_p + \ldots + n_1)-p} \, n_p \cdots n_1}{n_p (n_p + n_{p-1}) \cdots (n_p + \ldots + n_1)} \, 
\B_{n_1} \cdots \B_{n_p}
\Big\}
\, . \label{A+B:genB-3}
\end{align}
This is the explicit description
of (\ref{Zassenhaus-formula}) without using the functions $Z_n$.
We should notice that the ordering of the operators $\B_{n_i}$ is different from that of (\ref{A+B:genB-2}).

The descriptions
we obtained are quite useful if the product of the operator $\B_{n_i}$ is truncated at a certain level such as $\B_{n_k} \B_{n_{k-1}} \cdots \B_{n_1} = 0$,
which originates from the nilpotency of the operator $B$ of degree $k$, i.e., $B^k = 0$.

\subsection*{Baker-Campbell-Hausdorff formula}

We can also discuss the 
Baker-Campbell-Hausdorff (BCH, for short) formula
\begin{align}
\e^Z \ = \ \e^X \e^Y
\ &= \ 
\exp \Big\{
X + Y + \half [X,Y]
+ \frac{1}{12} \big( [X,[X,Y]] + [Y,[Y,X]] \big)
+ \ldots
\Big\}
\, , 
\label{BCH}
\end{align}
by using the descriptions (\ref{A+B:genB-2}) and (\ref{A+B:genB-3}), though
the general form (\ref{BCH}) 
has already been well-known 
(see, for instance, \cite{Varadarajan}).
We would like to use the operator  $\e^Z$ rather than $Z$,
because we often encounter the exponential form such as $\e^X \e^Y$ in quantum mechanics.
Multiplying (\ref{A+B:genB-2}) by $\e^{-A}$ from the right and replacing $A+B$ and $-A$ with $X$ and $Y$ respectively, we obtain
\bsubeq \label{BCH-another1}
\begin{align}
\e^X \e^Y
\ &= \ 
1 + \sum_{p=1}^{\infty} \sum_{n_1, \ldots, n_p=1}^{\infty} 
\frac{(-1)^{(n_p + \ldots + n_1) - p} \, n_p \cdots n_1}{n_p (n_p + n_{p-1}) \cdots (n_p + \ldots + n_1)} \, \X_{n_p} \cdots \X_{n_1}
\, , \\
\X_{n}
\ &\equiv \ 
\frac{1}{n!} (\mathscr{L}_{Y})^{n-1} (X+Y)
\, .
\end{align}
\esubeq
On the other hand, multiplying (\ref{A+B:genB-3}) by $\e^{-A}$ from the left and replacing $-A$ and $A+B$ with $X$ and $Y$ respectively, we find 
\bsubeq \label{BCH-another2}
\begin{align}
\e^X \e^Y
\ &= \ 
1 + \sum_{p=1}^{\infty} \sum_{n_1, \ldots, n_p=1}^{\infty} 
\frac{n_p \cdots n_1}{n_p (n_p + n_{p-1}) \cdots (n_p + \ldots + n_1)} \, 
\Y_{n_1} \cdots \Y_{n_p}
\, , \\
\Y_{n}
\ &\equiv \ 
\frac{1}{n!} (\mathscr{L}_{X})^{n-1} (X+Y)
\, . 
\end{align}
\esubeq
The original BCH formula $Z = \log (\e^X \e^Y)$ consists only of $X$, $Y$ and their commutators.
On the other hand, we immediately find that powers of $X$ and $Y$ directly contribute to the expansion in both (\ref{BCH-another1}) and (\ref{BCH-another2}).
These two seem to be different feature. 
However, the powers of $X$ and $Y$ in (\ref{BCH-another1}) and (\ref{BCH-another2}) originate from the corresponding power of $X+Y$ in the Taylor expansion of (\ref{BCH}).

We recognize that
the operators $X$ and $Y$ 
in (\ref{BCH-another1}) and (\ref{BCH-another2})
do not appear on equal footing with each other.
In order to describe an expression on equal footing, we simply sum up (\ref{BCH-another1}) and (\ref{BCH-another2}), and divide it by two.
For instance, we evaluate this up to cubic powers of the operators $X$ and $Y$ in such a way that
\begin{align*}
\half \big\{ &\text{(\ref{BCH-another1})} + \text{(\ref{BCH-another2})} \big\}
\nn \\
\ &= \ 
\half \Big\{
1 
+ \Big( \X_1 - \X_2 + \X_3 + \ldots \Big)
+ \Big(
\frac{1}{2} \X_1 \X_1
- \frac{1}{3} \X_2 \X_1
- \frac{2}{3} \X_1 \X_2
+ \ldots
\Big)
+ \Big(
\frac{1}{6} \X_1 \X_1 \X_1
+ \ldots
\Big)
\Big\}
\nn \\
\ & \ \ \ \ 
+ \half \Big\{
1 
+ \Big(
\Y_1 
+ \Y_2 
+ \Y_3
+ \ldots
\Big)
+ \Big(
\frac{1}{2} \Y_1 \Y_1
+ \frac{2}{3} \Y_2 \Y_1
+ \frac{1}{3} \Y_1 \Y_2
+ \ldots
\Big)
+ \Big(
\frac{1}{6} \Y_1 \Y_1 \Y_1
+ \ldots
\Big)
\Big\} 
\nn \\
\ &= \ 
1 
+ (X + Y)
- \frac{1}{4} \Big( [Y,X] - [X,Y] \Big) 
+ \frac{1}{12} \Big( [Y,[Y,X]] + [X,[X,Y]] \Big) 
\nn \\
\ & \ \ \ \ 
+ \frac{1}{2} (X + Y)^2
- \frac{1}{12} \Big( [Y,X] (X+Y) - (X+Y) [X,Y] \Big) 
- \frac{1}{6} \Big( (X+Y) [Y,X] - [X,Y] (X+Y) \Big) 
\nn \\
\ & \ \ \ \ 
+ \frac{1}{6} (X+Y)^3
+ \ldots
\nn \\
\ &= \ 
1 
+ (X + Y) 
+ \Big\{ \frac{1}{2} [X,Y] + \frac{1}{2} (X + Y)^2 \Big\}
\nn \\
\ & \ \ \ \
+ \Big\{ \frac{1}{12} \Big( [Y,[Y,X]] + [X,[X,Y]] \Big) 
+ \frac{1}{4} \Big( [X,Y] (X+Y) + (X+Y) [X,Y] \Big)
+ \frac{1}{3!} (X+Y)^3 \Big\}
+ \ldots
\, .
\end{align*}
This coincides with the Taylor expansion of $\e^X \e^Y$ in the form of (\ref{BCH}).
Indeed, we can find the coincidence in any powers of the operators.
Once we establish the above new descriptions,
it would be interesting to apply it to a generalization of the BCH formula such as $\e^X \e^Y \e^Z$ developed in \cite{Matone}.

\subsection*{Acknowledgments}

I would like to thank 
Reona Arai,
Tetsutaro Higaki,
Hideaki Iida,
Hiroyasu Miyazaki,
Toshifumi Noumi,
Noriaki Ogawa,
Masato Taki,
Akinori Tanaka
and 
Masahide Yamaguchi
for helpful discussions.
I would also like to thank
Katsushi Ito,
Marco Matone
and 
Hector Moya-Cessa
for valuable correspondence.
I am supported by the Iwanami-Fujukai Foundation.
I am also supported in part by the MEXT-Supported Program for the Strategic Research Foundation at Private Universities ``Topological Science'' ({Grant No.~S1511006}) and by the MEXT Grant-in-Aid for Scientific Research on Innovative Areas ``Nuclear Matter in Neutron Stars Investigated by Experiments and Astronomical Observations'' (No.~15H00841 by Muneto Nitta).

\subsection*{Appendix}

Here we explicitly exhibit a series of $X_{m,p}$ defined in (\ref{Xm-Xmp}).
When we consider $(A+B)^2$ as the form (\ref{A+B^n}), we obtain $X_2$ and $X_{2,p}$ as follows:
\begin{alignat*}{3}
X_2 
\ &= \ 
B^2 + \B'_2
\, , &\ls
X_{2,1}
\ &= \ 
\B'_2
\, , &\ls
X_{2,2}
\ &= \ 
B^2
\, .
\end{alignat*}
In the case of $(A+B)^3$, the components $X_{3,p}$ are
\begin{alignat*}{3}
X_{3,1}
\ &= \ 
\B'_3
\, , &\ls
X_{3,2}
\ &= \ 
\B'_2 B + 2 B \B'_2
\, , &\ls
X_{3,3}
\ &= \ 
B^3
\, .
\end{alignat*}
In the same way, the explicit forms of $X_{4,p}$ and $X_{5,p}$ are given as
\begin{align*}
X_{4,1}
\ &= \ 
\B'_4
\, , \\
X_{4,2}
\ &= \ 
\B'_3 B + 3 (\B'_2)^2 + 3 B \B'_3
\, , \\
X_{4,3}
\ &= \
\big( \B'_2 B + 2 B \B'_2 \big) B
+ 3 B^2 \B'_2
\, , \\
X_{4,4}
\ &= \ 
B^4
\, , \\
X_{5,1}
\ &= \ 
\B'_5
\, , \\
X_{5,2}
\ &= \
\B'_4 B
+ 4 \B'_3 \B'_2
+ 6 \B'_2 \B'_3
+ 4 B \B'_4
\, , \\
X_{5,3}
\ &= \
\big( \B'_3 B + 3 (\B'_2)^2 + 3 B \B'_3 \big) B
+ 4 \big( \B'_2 B + 2 B \B'_2 \big) \B'_2
+ 6 B^2 \B'_3
\, , \\
X_{5,4}
\ &= \
\big\{ \big( \B'_2 B + 2 B \B'_2 \big) B + 3 B^2 \B'_2 \big\} B
+ 4 B^3 \B'_2 
\, , \\
X_{5,5}
\ &= \ 
B^5
\, .
\end{align*}

Here we prove (\ref{sol:Xmp-gen}) as the solution of the recursion relation (\ref{re:Xmp-gen}) by mathematical induction. 
Let us assume that each $X_{k,l}$ with $1 \leq l \leq k \leq m$ satisfies the expression (\ref{sol:Xmp-gen}).
We compute $\mathscr{L}_A X_{m,p} + B X_{m,p-1}$:
\begin{align*}
\mathscr{L}_{A} &X_{m,p}
+ B X_{m,p-1}
\nn \\
\ &= \ 
\sum_{k=1}^{m-p+1} \frac{(m-1)!}{(k-1)! (m-k)!} 
\Big\{ (\mathscr{L}_{A} X_{m-k,p-1}) \B'_k + X_{m-k,p-1} \B'_{k+1} \Big\}
+ \sum_{k=1}^{m-p+2} \frac{(m-1)!}{(k-1)! (m-k)!} 
\, B X_{m-k,p-2} \B'_k 
\nn \\
\ &= \ 
\sum_{k=1}^{m-p+1} \frac{(m-1)!}{(k-1)! (m-k)!} 
\Big\{ \big( X_{m+1-k,p-1} - B X_{m-k,p-2} \big) \B'_k + X_{m-k,p-1} \B'_{k+1} \Big\}
\nn \\
\ & \ \ \ \
+ \sum_{k=1}^{m-p+2} \frac{(m-1)!}{(k-1)! (m-k)!} 
\, B X_{m-k,p-2} \B'_k 
\nn \\
\ &= \ 
\sum_{k=1}^{m-p+1} \frac{(m-1)!}{(k-1)! (m-k)!} 
\Big\{ X_{m+1-k,p-1} \B'_k + X_{m-k,p-1} \B'_{k+1} \Big\}
+ B X_{p-2,p-2} \B'_{m-p+2}
\nn \\
\ &= \ 
\sum_{k=1}^{(m+1)-p+1} \frac{m!}{(k-1)! (m+1-k)!} 
\, X_{(m+1)-k,p-1} \B'_k
\, . 
\end{align*}
The final form is nothing but $X_{m+1,p}$.
Hence we proved that
(\ref{sol:Xmp-gen}) is the solution of the relation (\ref{re:Xmp-gen}).


}
\end{document}